\documentclass[11pt]{article}

\usepackage[preprint]{acl}

\usepackage{times}
\usepackage{latexsym}
\usepackage[T1]{fontenc}
\usepackage[utf8]{inputenc}
\usepackage{microtype}
\usepackage{inconsolata}
\usepackage{amsmath,amssymb,graphicx,booktabs,array}
\usepackage{url}

\newcommand{\ours}{CLRM}

\definecolor{aclblue}{rgb}{0,0,0.5}
\hypersetup{
    colorlinks=true,
    citecolor=aclblue,
    linkcolor=aclblue,
    urlcolor=aclblue,
    pdfauthor={Yangyang Qu, Siqing Qin, Massimiliano Todisco, Kong Aik Lee},
    pdftitle={CLRM: Cross-Layer Relation Modeling for Audio Deepfake Detection}
}

\title{CLRM: Cross-Layer Relation Modeling for Audio Deepfake Detection}

\author{
    Yangyang Qu \\
    EURECOM \\
    Sophia Antipolis, France
    \And
    Siqing Qin \\
    The Hong Kong Polytechnic University \\
    Hong Kong SAR, China
    \AND
    Kong Aik Lee \\
    The Hong Kong Polytechnic University \\
    Hong Kong SAR, China
    \And
    Massimiliano Todisco \\
    EURECOM \\
    Sophia Antipolis, France
}

\begin{document}
\maketitle
\begin{abstract}
Audio deepfake detection (ADD) aims to distinguish bonafide speech from synthetic or manipulated speech.
Representations from different layers of self-supervised learning (SSL) speech encoders vary in their ability to distinguish bonafide from spoofed speech.
Comparing representations of the same utterance across encoder depths may reveal useful spoofing cues, yet common layer-selection and aggregation methods do not provide these comparisons as detection inputs.
We propose Cross-Layer Relation Modeling (\ours{}), which explicitly encodes cross-layer relations alongside layer-specific features, making both directly available for ADD.
\ours{} reuses the existing detector's SSL encoder and identifies informative layer pairs by combining speaker-relative representation geometry with source-domain detection performance.
The resulting relation score is fused with the detector's original score, requiring only the training of compact relation heads and fusion parameters while keeping the existing detector frozen.
Across five external datasets, \ours{} reduces the mean equal error rate (EER) for all three evaluated detectors, achieving relative reductions of 7.8--16.7\%.
Controlled comparisons with identical layer summaries and approximately matched auxiliary-head parameter counts also show a lower mean EER for explicit relations than direct concatenation, both before and after score fusion.
These findings support explicit cross-layer comparisons as useful detection inputs for SSL-based audio deepfake detectors.
\end{abstract}

\section{Introduction}
\label{sec:introduction}

Audio deepfake detection (ADD) aims to distinguish bonafide speech from synthetic or manipulated speech.
Detecting spoofed speech across different generation methods and recording conditions remains an important challenge \cite{muller2022does,huang-etal-2026-data}.
Pretrained self-supervised learning (SSL) speech encoders, such as the multilingual XLS-R model \cite{babu2021xls}, are widely used for ADD.
These encoders produce representations at multiple processing depths, and previous analyses have shown that representations from different layers vary in their ability to distinguish bonafide from spoofed speech \cite{elkheir2025layerwise}.
To exploit this variation, existing detectors select informative layers, aggregate features from multiple depths, or combine layer-specific predictions \cite{zhang2024xlsrsls,pan2024attm,xiao2025decisionfusion}.
For example, XLSR-SLS employs a Sensitive Layer Selection (SLS) module to aggregate XLS-R representations using learned layer-dependent weights \cite{zhang2024xlsrsls}.

These approaches demonstrate the value of multilayer features for ADD, motivating us to investigate whether explicit comparisons between layers provide useful detection cues beyond those learned from feature concatenation.
Representations at different encoder depths offer distinct views of the same utterance, making their differences and interactions potentially informative for detection.
Direct concatenation preserves these representations but leaves the classifier to learn how to compare them.
We therefore construct explicit cross-layer comparisons alongside layer-specific features, rather than relying on the classifier to learn these comparisons from direct concatenation.

In this paper, we propose Cross-Layer Relation Modeling (\ours{}), a detector-augmentation framework that explicitly models relations between selected SSL layer representations.
For each utterance, \ours{} reuses the existing detector's SSL encoder and constructs a joint representation from two selected layers, which retains their individual features while encoding their pairwise discrepancies, interactions, and similarity.
Compact relation heads learn an auxiliary detection score from this representation, which is combined with the original detector's score through score-level fusion.
The existing detector, including its SSL encoder, remains frozen, and training is confined to the relation heads and fusion parameters.
This design enables existing SSL-based detectors to incorporate relation-based detection evidence without modifying their architectures or retraining their original parameters.

The choice of layer pair determines which cross-layer relationships are modeled by CLRM.
We use bonafide speech to summarize speaker-relative organization at different encoder depths, prioritizing candidate pairs that are separated in depth but have similar aggregate speaker statistics.
Detection error on held-out source attacks determines the final layer pair and fusion configuration.
The selected configuration is then applied unchanged to external datasets.

We evaluate \ours{} with three XLS-R-based detector architectures on five external datasets.
The five-dataset mean equal error rate (EER) decreases for all three detectors, with a maximum relative reduction of 16.7\%.
With identical layer summaries and approximately matched auxiliary-head parameter counts, explicit relation features also achieve lower mean EER than direct concatenation before and after fusion.

Our contributions are threefold:
\begin{itemize}
    \item \textbf{Explicit cross-layer relation representation.}
    We introduce a detection input that retains selected SSL layer summaries and directly encodes their discrepancies and interactions.

    \item \textbf{Source-guided detector augmentation.}
    We combine speaker-relative layer-pair candidate construction, source-based configuration selection, and relation-score fusion to augment frozen XLS-R-based detectors with compact trainable heads.

    \item \textbf{Controlled empirical evaluation.}
    We demonstrate improvements across three detector architectures and five external datasets.
    With identical layer summaries and approximately matched auxiliary-head parameter counts, explicit relations achieve a lower five-dataset mean EER than direct concatenation.
\end{itemize}

\begin{figure*}[!t]
\centering
\includegraphics[width=\textwidth]{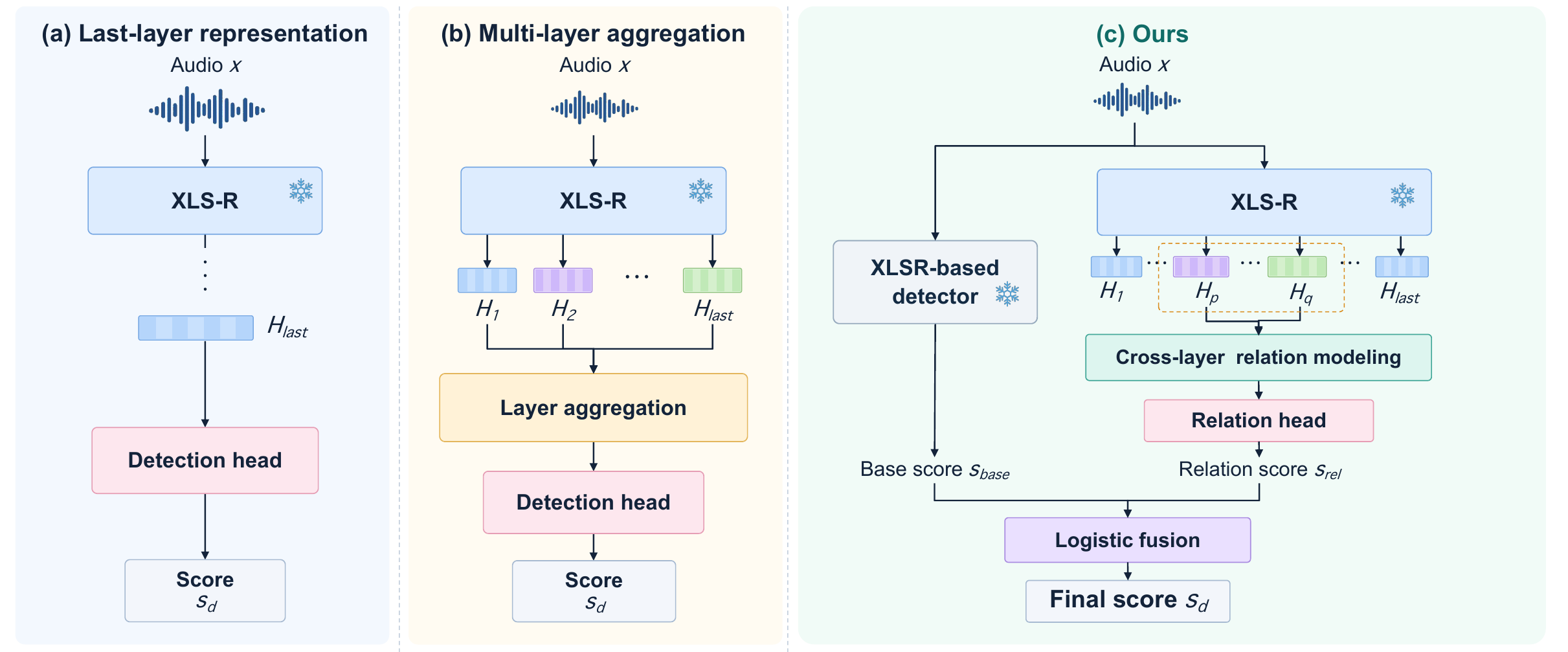}
\caption{
\textbf{CLRM for ADD.}
(a) Last-layer detection and (b) multilayer aggregation illustrate conventional uses of SSL representations.
(c) \ours{} uses selected representations from the frozen XLS-R encoder of the existing detector and combines the resulting relation score with its original score through source-fitted fusion.
Snowflakes indicate frozen components.
}
\label{fig:model}
\end{figure*}

\section{Related Work}
\label{sec:related_work}

\subsection{Multilayer SSL Representations for ADD}
Pretrained SSL speech encoders provide representations at multiple depths that are widely used for ADD.
\citet{elkheir2025layerwise} analyze the detection capabilities of different SSL layers, revealing substantial variation across encoder depths.
Several methods exploit these representations through layer selection and aggregation.
XLSR-SLS \cite{zhang2024xlsrsls} learns layer-dependent weights to aggregate XLS-R features, while \citet{pan2024attm} introduce attentive merging of intermediate WavLM representations.
\citet{xiao2025decisionfusion} instead combine predictions from individual XLS-R layers.
These studies demonstrate the effectiveness of leveraging multilayer information through feature aggregation and decision fusion.

\subsection{Cross-Layer Dependencies and Representation Comparisons}
Beyond feature aggregation, recent studies have explored relationships between representations.
HierCon \cite{liang2026hiercon} employs hierarchical attention and contrastive learning to model dependencies across temporal frames and encoder layers.
Explicit feature comparisons have also been investigated in representation learning.
Sentence-BERT \cite{reimers-gurevych-2019-sentence} explores pairwise embedding combinations involving elementwise differences and products for sentence-pair classification.
These studies motivate examining explicit comparisons between selected SSL layer representations as detection inputs.

\subsection{Cross-Dataset ADD}
Generalization across synthesis methods and recording conditions remains a major challenge in ADD \cite{muller2022does}.
\citet{li2024crossdomain} investigate cross-domain detection using speech generated by zero-shot text-to-speech systems.
SpeechFake \cite{huang-etal-2025-speechfake} introduces a large-scale multilingual dataset covering diverse speech generation methods.
\citet{huang-etal-2026-data} investigate source and generator diversity to improve detection generalization.
ICLAD \cite{chou-etal-2026-iclad} explores comparison-guided in-context learning with audio language models for detecting unseen deepfakes.
These studies address cross-dataset detection through data diversity, evaluation benchmarks, and alternative detection strategies.

\section{Cross-Layer Relation Modeling}
\label{sec:method}

\subsection{Framework Overview}

For an input utterance $x$, \ours{} augments a frozen SSL-based detector with a learned score derived from cross-layer relations.
The existing detector produces $s^{\mathrm{base}}(x)$, and its SSL encoder supplies the hidden representations used to construct a relation input from selected layers $(p,q)$.
Relation heads map this input to $s^{\mathrm{rel}}_{p,q}(x)$, which is fused with the original score to obtain $s^{\mathrm{final}}(x)$.
Source-side geometry guides candidate layer-pair construction, followed by detection-based configuration selection.
The detector and encoder remain frozen; only relation heads and fusion parameters are fitted.

\subsection{Explicit Cross-Layer Relation Representation}

Our representation is designed to retain information from individual encoder layers while explicitly describing how their features differ and interact.
We first summarize the hidden states at each layer into a fixed-dimensional utterance representation and then construct pairwise comparisons between the selected summaries.
Let $\mathbf{H}^{(l)}(x)\in\mathbb{R}^{T^{(l)}(x)\times d}$ denote the output of Transformer block $l$, where the 24 blocks are indexed as L1--L24 and $T^{(l)}(x)$ and $d$ denote the frame count and hidden dimension.
For each layer, we concatenate the temporal mean $\boldsymbol{\mu}_l(x)$ and population standard deviation $\boldsymbol{\sigma}_l(x)$:
\begin{equation}
\mathbf{z}_l(x)
=[\boldsymbol{\mu}_l(x);\boldsymbol{\sigma}_l(x)]
\in\mathbb{R}^{2d}.
\label{eq:layer-summary}
\end{equation}
Here, $[\cdot;\cdot]$ denotes concatenation.
Each coordinate of $\mathbf{z}_l(x)$ is standardized using statistics estimated for that layer from the official training split of ASVspoof 2019 Logical Access (LA), yielding $\tilde{\mathbf{z}}_l(x)$.
The same standardization is applied to development and evaluation utterances.

For the selected layers $p$ and $q$, we construct the relation representation as
\begin{equation}
\begin{aligned}
\mathbf{r}_{p,q}(x)=[&
\tilde{\mathbf{z}}_p;\tilde{\mathbf{z}}_q;
|\tilde{\mathbf{z}}_p-\tilde{\mathbf{z}}_q|;\\
&\tilde{\mathbf{z}}_p\odot\tilde{\mathbf{z}}_q;
\cos(\tilde{\mathbf{z}}_p,\tilde{\mathbf{z}}_q)].
\end{aligned}
\label{eq:cross-layer-relation}
\end{equation}
The argument $x$ is omitted from the layer summaries on the right-hand side for readability.
The operators $|\cdot|$, $\odot$, and $\cos(\cdot,\cdot)$ denote elementwise absolute value, elementwise multiplication, and cosine similarity, respectively.
The absolute difference represents coordinate-wise discrepancies, the elementwise product encodes multiplicative interactions, and cosine similarity summarizes the angular relationship between the two summaries.
Together with the retained layer summaries, these comparisons form an $(8d+1)$-dimensional input that makes both layer-specific and relational information available to the classifier.
Since the comparisons operate on utterance-level summaries, no frame-wise alignment between the selected layers is required.

\subsection{Relation Scoring}

We learn a detection score from $\mathbf{r}_{p,q}(x)$ using three independently initialized feed-forward heads.
Each head $g_{\phi_m}$ consists of a linear projection, a Gaussian error linear unit (GELU) activation, dropout, and a linear output layer.
The relation score is obtained by averaging the raw outputs of the heads:
\begin{equation}
s^{\mathrm{rel}}_{p,q}(x)
=\frac{1}{M}\sum_{m=1}^{M}
g_{\phi_m}(\mathbf{r}_{p,q}(x)),
\qquad M=3.
\label{eq:relation-score}
\end{equation}
The heads are fitted independently on the official source training split, and their checkpoints are selected using the official source development split.
Score averaging is performed before orientation, normalization, and fusion.
For XLSR-SLS, we select the sign of the relation score on Meta-dev, a source partition used for configuration selection.
The selection criterion is detailed in Appendix~\ref{app:fitting}.

\subsection{Score-Level Fusion}

We combine the relation score with the score of the existing detector to integrate cross-layer information into its final decision.
For a fixed selected pair $(p,q)$, we write $s^{\mathrm{rel}}(x)=s^{\mathrm{rel}}_{p,q}(x)$.
The original and relation scores are standardized using their respective statistics from the official source development split, yielding $\tilde{s}^{\mathrm{base}}(x)$ and $\tilde{s}^{\mathrm{rel}}(x)$.
We then apply logistic-regression score fusion:
\begin{equation}
s^{\mathrm{final}}(x)
=a\tilde{s}^{\mathrm{base}}(x)
+b\tilde{s}^{\mathrm{rel}}(x)+c,
\label{eq:score-fusion}
\end{equation}
where $a$, $b$, and $c$ are fusion coefficients fitted on the official source development split.
The fused score is the linear decision score of the logistic-regression model, with its regularization strength selected on Meta-dev.
This score-level interface retains the existing detector's feature processing and decision pathway while incorporating the relation-based score.
The SSL encoder and original detector remain frozen throughout training; only the relation heads and fusion parameters are optimized.

\begin{table*}[t]
\centering
\small
\setlength{\tabcolsep}{5pt}
\begin{tabular*}{\textwidth}{@{\extracolsep{\fill}}llrrc@{}}
\toprule
Protocol & Geometry reference & Meta-dev $N$ & Meta-test $N$ & Overlap U/S \\
\midrule
Attack-disjoint & Eval bonafide: 7,355 utt., 67 spk. & 41,753 & 36,839 & 7,355/67 \\
Utterance-disjoint & Train bonafide: 2,580 utt., 20 spk. & 38,085 & 33,152 & 0/67 \\
Speaker-disjoint & Train bonafide: 2,580 utt., 20 spk. & 20,718 & 18,605 & 0/0 \\
\bottomrule
\end{tabular*}
\caption{\textbf{Source evaluation protocols.}
$N$ is the number of trials, and U/S denotes bonafide utterance/speaker overlap between Meta-dev and Meta-test.
Meta-dev is used for configuration selection and Meta-test for assessment.}
\label{tab:source-protocols}
\end{table*}

\subsection{Geometry-Guided Layer-Pair Selection}

The choice of layer pair determines which cross-layer relationships are presented to the relation classifier.
We construct a compact candidate set by favoring pairs that are separated in encoder depth but exhibit similar speaker-relative organization on bonafide speech.
Depth separation encourages comparisons between different processing stages, while speaker-relative statistics provide a common reference for comparing layers.
Detection performance is then used to select the final pair from these candidates.

For each layer $l$, we characterize speaker-relative organization using bonafide source utterances with speaker labels.
For candidate selection, we obtain geometry embeddings through temporal mean pooling of $\mathbf{H}^{(l)}(x)$ followed by $\ell_2$ normalization. These embeddings differ from the standardized mean-and-standard-deviation summaries used to construct the relation representation in Section~3.2.
Each embedding is compared with its own-speaker centroid and competing-speaker centroids, using a leave-one-out centroid for the correct speaker.
We aggregate these comparisons into four layer-level statistics:
\begin{equation}
\mathbf{g}_l=[R_l;A_l;M_l^{\max};M_l^{\mathrm{mean}}],
\label{eq:geometry-summary}
\end{equation}
where $R_l$ is the mean rank of the correct speaker, $A_l$ denotes top-1 own-speaker identification accuracy, and $M_l^{\max}$ and $M_l^{\mathrm{mean}}$ are the mean similarity margins relative to the strongest and average competing speakers, respectively.
Their precise definitions are provided in Appendix~\ref{app:geometry-statistics}.

\begin{figure}[t]
\centering
\includegraphics[width=\columnwidth]{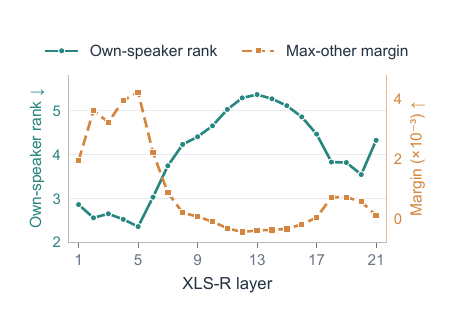}
\caption{\textbf{Speaker-relative statistics across XLS-R layers.}
Own-speaker rank (lower is better) and max-other margin (higher is better) are two of the four statistics used for candidate construction.
Both are computed from bonafide ASVspoof 2019 LA evaluation utterances under the primary source protocol.
The figure focuses on L1--L21, as the substantially different scales of L22--L24 would make variations within the candidate range less visible.}
\label{fig:geometry}
\end{figure}

Figure~\ref{fig:geometry} illustrates how speaker-relative organization varies with encoder depth.
In the shallower layers, own-speaker rank is relatively low and max-other margin is relatively high; toward the middle layers, rank increases while the margin decreases, followed by further changes at deeper layers.
These non-monotonic patterns motivate comparing layers using their speaker-relative summaries rather than relying on depth alone.
We use all four statistics in $\mathbf{g}_l$ to construct the candidate ranking.

After standardizing each component of $\mathbf{g}_l$ across candidate layers to obtain $\hat{\mathbf{g}}_l$, we define the geometry distance and ranking score for layers $p$ and $q$ as
\begin{equation}
\begin{aligned}
D_G(p,q)&=\|\hat{\mathbf{g}}_p-\hat{\mathbf{g}}_q\|_2,\\
R_G(p,q)&=\frac{|q-p|}{D_G(p,q)+\epsilon},
\end{aligned}
\label{eq:geometry-ranking}
\end{equation}
where $\epsilon=10^{-8}$.
A higher $R_G(p,q)$ favors greater separation in encoder depth together with more similar speaker-relative summaries.
We consider layers L1--L21 with a minimum separation of eight layers, yielding 91 eligible pairs, and retain the six highest-ranked candidates.
The geometry statistics are used for candidate construction and do not enter the utterance-level relation representation in Eq.~\ref{eq:cross-layer-relation}.

For each candidate, the relation heads are trained on the official source training split and fusion coefficients are fitted on the official development split.
We select the layer pair and inverse fusion-regularization strength $C$ on \emph{Meta-dev} by minimizing fused Macro equal error rate (EER), defined as the unweighted mean of attack-wise EERs.
The selected configuration is evaluated on \emph{Meta-test}, which contains different spoof attacks, and then applied unchanged to the external datasets.
Section~\ref{sec:data-protocols} specifies the source partitions and geometry-reference populations.
Speaker labels and geometry statistics are not required at inference.

\section{Experimental Setup}
\label{sec:experiments}

\subsection{Data and Evaluation Protocols}
\label{sec:data-protocols}

We use ASVspoof 2019 Logical Access (LA) for source-side development \cite{wang2020asvspoof2019}.
Its official training split is used to fit the relation heads and estimate layer-wise feature-standardization statistics, while the official development split is used for checkpoint selection, score normalization, and fusion-parameter fitting.

We partition the official evaluation split by spoof attack into \emph{Meta-dev} for configuration selection and \emph{Meta-test} for assessing the selected configuration.
The primary protocol shares bona fide utterances between these partitions and uses them as the geometry reference.
We additionally evaluate utterance-disjoint and speaker-disjoint protocols, reconstructing geometry from bona fide training data and repeating configuration selection.
Table~\ref{tab:source-protocols} summarizes these protocols, with further details in Appendix~\ref{app:protocol-construction}.

For cross-dataset evaluation, we use ASVspoof 2021 LA and Deepfake (DF) \cite{liu2023asvspoof2021}, ASVspoof 5 Eval \cite{wang2024asvspoof5}, the normalized Fake-or-Real test set \cite{reimao2019for}, and In-the-Wild \cite{muller2022does}.
The layer pair, relation heads, score orientation, normalization statistics, and fusion parameters are determined from the source data and held fixed during external evaluation.
The ASVspoof 2021 results follow this post-challenge source-development protocol.\footnote{Our source configuration selection uses the ASVspoof 2019 LA evaluation partition, which was excluded from system development under the original ASVspoof 2021 challenge rules \cite{liu2023asvspoof2021}.}
Dataset-specific evaluation procedures are described in Appendix~\ref{app:reproducibility}.

\subsection{Evaluated Detectors}
\label{sec:evaluated-detectors}

We evaluate \ours{} with three detector architectures using XLS-R representations: XLSR-SLS \cite{zhang2024xlsrsls}, XLSR-Conformer-TCM based on Temporal-Channel Modeling (TCM) \cite{truong24b_interspeech}, and XLSR-AASIST, an XLS-R-based variant of Audio Anti-Spoofing using Integrated Spectro-Temporal Graph Attention Networks (AASIST) \cite{jung2022aasist,tak2022automatic}.
XLSR-SLS is the primary detector for studying the relation representation and selecting the layer pair.
The selected layer indices are reused for XLSR-Conformer-TCM and XLSR-AASIST, with relation heads and fusion parameters fitted separately for each detector using source data.
For each augmented system, \ours{} uses representations from the detector's existing frozen XLS-R encoder.
Each base detector and its augmentation are evaluated on the same external trials using the same dataset-specific evaluator.
Additional SSL-based reference results are included in Table~\ref{tab:main_results}, with their result sources summarized in Appendix~\ref{app:result-provenance}.

\subsection{Controlled Representation Comparisons}
\label{sec:representation-controls}

To isolate the effect of explicit cross-layer relations, we compare the proposed \emph{Relation} input with direct concatenation of the same selected layer summaries.
\emph{Relation} denotes the representation in Eq.~\ref{eq:cross-layer-relation}, whereas \ours{} denotes the complete augmentation with relation scoring and fusion.
All representation controls use the selected L8--L21 layers and identical layer-wise standardization statistics.
The concatenation input is
\begin{equation}
\mathbf{r}^{\mathrm{concat}}_{p,q}(x)
=[\tilde{\mathbf{z}}_p(x);\tilde{\mathbf{z}}_q(x)]
\in\mathbb{R}^{4d}.
\label{eq:concat-control}
\end{equation}
We evaluate Concat-256, which matches the hidden width of the relation head, and Concat-512, which approximately matches its trainable parameter count.
Each input configuration uses three independently fitted heads with the same source-side training and selection procedure.

We assess the auxiliary scores both independently and after fusion with XLSR-SLS.
These comparisons measure the discriminative value of the input representation and its contribution to the existing detector, respectively.
Additional source-side controls examine single-layer inputs, individual relation terms, and alternative score-combination rules; their configurations and parameter counts are provided in Appendix~\ref{app:input-controls}.

\begin{table*}[t]
\centering
\small
\renewcommand{\arraystretch}{1.02}
\setlength{\tabcolsep}{2pt}
\begin{tabular}{@{}lrrrrrr@{}}
\toprule
\textbf{System}
& \textbf{21 LA}
& \textbf{21 DF}
& \textbf{ASVspoof 5}
& \textbf{In-the-Wild}
& \textbf{Fake-or-Real}
& \textbf{Average} \\
\midrule
\multicolumn{7}{@{}l}{\emph{Reference systems}} \\
XLSR-Conformer \cite{rosello2023conformer}
& 1.28 & 2.86 & 17.21 & 10.88 & 10.03 & 8.45 \\
Nes2Net-X \cite{liu2025nes2net}
& 2.00 & 1.78 & 22.06 & 6.60 & 6.31 & 7.75 \\
XLSR-DADGMoE-AASIST \cite{qin2026dadgmoe}
& 1.53 & 2.54 & 18.01 & \textbf{6.35} & 4.42 & 6.57 \\
BiCrossMamba-ST \cite{elkheir2025bicrossmamba}
& 3.83 & 2.35 & \textbf{13.67} & 7.94 & 6.85 & 6.93 \\
\midrule
\multicolumn{7}{@{}l}{\emph{Base detectors and CLRM augmentation}} \\
XLSR-SLS \cite{zhang2024xlsrsls}
& 2.95 & \textbf{1.77} & 18.76 & 7.28 & 5.03 & 7.16 \\
\quad + \ours{}
& 2.56 & 1.91 & 16.32 & 7.45 & \textbf{4.08} & 6.46 \\
\addlinespace[2pt]
XLSR-Conformer-TCM \cite{truong24b_interspeech}
& 2.30 & 2.06 & 16.16 & 7.27 & 6.47 & 6.85 \\
\quad + \ours{}
& 2.23 & 2.06 & 14.55 & 7.13 & 5.61 & 6.32 \\
\addlinespace[2pt]
XLSR-AASIST \cite{jung2022aasist}
& 1.27 & 3.46 & 15.27 & 8.08 & 7.77 & 7.17 \\
\quad + \ours{}
& \textbf{1.17} & 2.92 & 14.41 & 6.70 & 4.66 & \textbf{5.97} \\
\bottomrule
\end{tabular}
\caption{\textbf{Detection performance on five external datasets} (EER, \%; lower is better).
The lower block compares each base detector with its \ours{} augmentation on the same evaluation trials and using the same evaluators.
21 LA and 21 DF denote ASVspoof 2021 LA and DF, respectively.
Average is the unweighted mean of the five dataset-level EERs (External-5); paired comparisons use unrounded values.
Reference-system result records are described in Appendix~\ref{app:result-provenance}.
Bold indicates the lowest value in each column.}
\label{tab:main_results}
\end{table*}

\subsection{Implementation and Evaluation Metrics}
\label{sec:implementation}

\paragraph{Implementation.}
We use the frozen XLS-R-300M encoder already employed by each base detector, with hidden dimension $d=1024$ \cite{babu2021xls}.
The relation representation has $8d+1=8{,}193$ dimensions.
Each relation head uses a hidden width of 256, GELU activation, and dropout of 0.20, yielding 2,097,921 trainable parameters per head and approximately 6.29 million across three heads.
The pretrained encoder and original detector remain frozen, and only the relation heads and fusion parameters are fitted.
Three independently initialized heads are trained for each configuration, and their raw outputs are averaged before score orientation, normalization, and fusion.
For the primary representation experiments, we use AdamW with learning rate $2\times10^{-4}$, weight decay $10^{-4}$, batch size 2,048, and a maximum of 15 epochs.
Training uses weighted binary cross-entropy and early stopping based on official-development EER with patience of four.
Audio is converted to mono at 16 kHz and truncated or repeated to 64,600 samples.
Appendix~\ref{app:optimization} provides the preprocessing, optimization, and seed settings.

\paragraph{Evaluation metrics.}
We report equal error rate (EER, \%), with lower values indicating better detection.
For source evaluation, \emph{Macro EER} averages attack-wise EERs, \emph{Worst EER} is the highest attack-wise EER, and \emph{Pooled EER} is computed over all trials in a partition.
For external evaluation, \emph{External-5} is the unweighted mean of the five dataset-level EERs.
EER is computed from the corresponding base, auxiliary-only, or fused score stream.
For the relation ensemble, raw head scores are averaged before evaluation.
ASVspoof 2021 LA and DF use the recorded official-evaluator results, while the remaining evaluation procedures are detailed in Appendix~\ref{app:reproducibility}.
For the source-side layer-pair diagnostic, we additionally report $\Delta_{\cos}=\bar{c}_{\mathrm{bona\ fide}}-\bar{c}_{\mathrm{spoof}}$, the difference in class-mean cosine similarity between the standardized layer summaries on the Meta-test.

\section{Results and Analysis}
\label{sec:results}

\subsection{Main Results across Detectors}
\label{sec:results-main}

Table~\ref{tab:main_results} shows that \ours{} lowers External-5 mean EER for all three detectors, with relative reductions of 9.7\%, 7.8\%, and 16.7\% for XLSR-SLS, XLSR-Conformer-TCM, and XLSR-AASIST, respectively.
The largest reduction is obtained with XLSR-AASIST (7.170\% to 5.972\%), which improves on all five datasets.
Together, the paired comparisons demonstrate the benefit of relation scoring across three XLS-R-based detector architectures.

\subsection{Effectiveness of Explicit Cross-Layer Relations}
\label{sec:results-relations}

\begin{table}[t]
\centering
\footnotesize
\setlength{\tabcolsep}{2.2pt}
\begin{tabular}{@{}lrrrr@{}}
\toprule
Auxiliary input
& \shortstack{Params/head\\($10^6$)}
& \shortstack{Source\\fused}
& \shortstack{Auxiliary\\only}
& \shortstack{Fused with\\SLS} \\
\midrule
None (SLS) & -- & 0.149 & -- & 7.158 \\
Concat-256 & 1.049 & 0.114 & 16.949 & 7.422 \\
Concat-512 & 2.098 & 0.117 & 16.728 & 7.330 \\
Relation & 2.098 & \textbf{0.069} & \textbf{11.333} & \textbf{6.462} \\
\bottomrule
\end{tabular}
\caption{\textbf{Explicit relations versus direct concatenation} (EER, \%).
All auxiliary inputs use identical L8--L21 layer summaries and standardization statistics.
Concat-256 matches the relation-head hidden width, and Concat-512 approximately matches its per-head parameter count.
Source fused is Meta-test Macro EER; the last two columns report External-5 EER before and after fusion with XLSR-SLS.}
\label{tab:relation_controls}
\end{table}

\begin{table*}[t]
\centering
\small
\setlength{\tabcolsep}{5pt}
\begin{tabular}{@{}lrrrrrr@{}}
\toprule
& \multicolumn{3}{c}{Fused with SLS}
& \multicolumn{3}{c}{Auxiliary only} \\
\cmidrule(lr){2-4}
\cmidrule(l){5-7}
Dataset
& Concat-256 & Concat-512 & Relation
& Concat-256 & Concat-512 & Relation \\
\midrule
2021 LA
& 2.301 & \textbf{2.288} & 2.558
& 4.799 & \textbf{4.712} & 6.081 \\
2021 DF
& 2.031 & 2.038 & \textbf{1.909}
& 8.165 & 8.319 & \textbf{5.416} \\
ASVspoof 5
& 19.905 & 19.374 & \textbf{16.317}
& 27.861 & 25.933 & \textbf{17.598} \\
In-the-Wild
& 8.580 & 8.548 & \textbf{7.450}
& 32.289 & 32.266 & \textbf{21.806} \\
Fake-or-Real
& 4.294 & 4.402 & \textbf{4.078}
& 11.632 & 12.408 & \textbf{5.762} \\
\midrule
Mean
& 7.422 & 7.330 & \textbf{6.462}
& 16.949 & 16.728 & \textbf{11.333} \\
\bottomrule
\end{tabular}
\caption{\textbf{Dataset-level comparison of relation and concatenation inputs} (EER, \%).
Each configuration uses the same L8--L21 summaries; Mean is the unweighted average across the five external datasets.
Bold marks the minimum within each dataset and scoring block.}
\label{tab:external_representation}
\end{table*}

\paragraph{Controlled representation comparison.}
With identical L8--L21 summaries, Relation achieves lower source Meta-test and External-5 mean EER than both concatenation controls (Table~\ref{tab:relation_controls}).
Against the near-parameter-matched Concat-512, source Macro EER decreases from 0.117\% to 0.069\%, while External-5 EER decreases from 16.728\% to 11.333\% before fusion and from 7.330\% to 6.462\% after fusion.
These controlled results show that explicit cross-layer comparisons provide more effective detection inputs than direct concatenation for the selected layers and classifier configurations.

\paragraph{Cross-dataset comparison.}
Relation yields the lowest EER among the tested representations on four of the five datasets, both before and after fusion (Table~\ref{tab:external_representation}).
On ASVspoof 5, for example, fused EER decreases from 19.374\% for Concat-512 to 16.317\% for Relation, with a corresponding improvement in auxiliary-only scoring.
The consistent gains across these two scoring conditions strengthen the evidence for relation-based inputs.

\begin{figure}[t]
\centering
\includegraphics[width=\columnwidth]{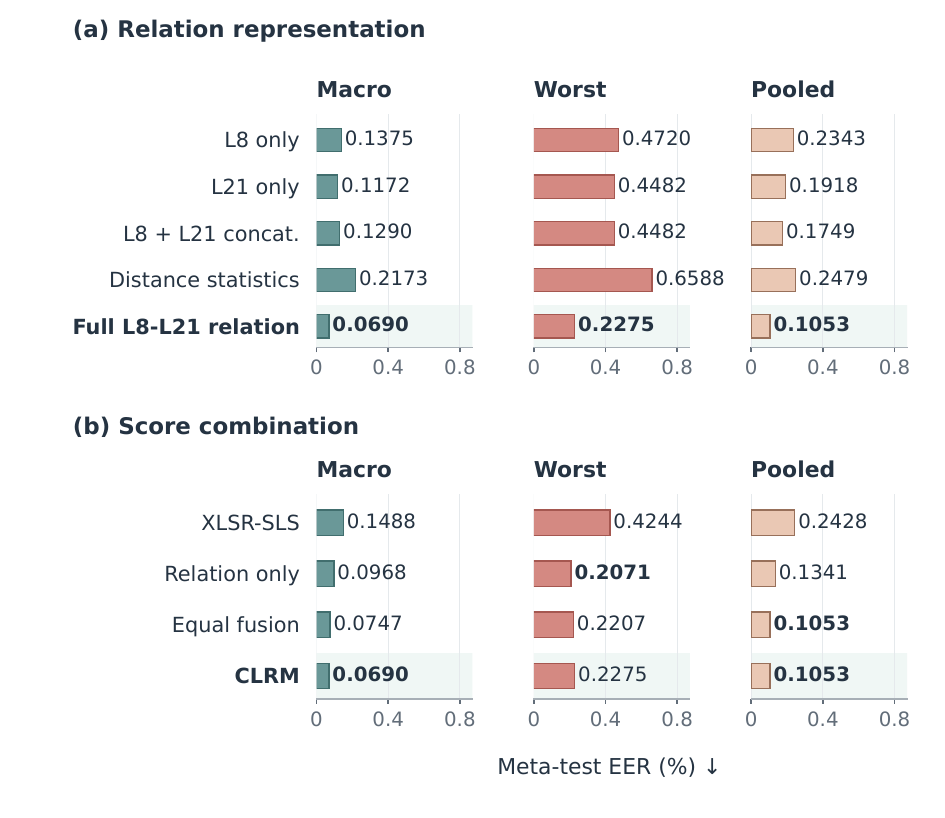}
\caption{\textbf{Source-side representation and score-combination ablations} (Meta-test EER, \%).
(a) Broad representation controls with their respective input dimensions and head capacities.
(b) XLSR-SLS, relation-only scoring, equal-weight fusion, and source-fitted logistic fusion.
Distance statistics comprise eight comparison measures detailed in Appendix~\ref{app:input-controls}.
The broad controls and the capacity-controlled experiments in Table~\ref{tab:relation_controls} are independently fitted configurations.}
\label{fig:relation}
\end{figure}

\paragraph{Representation components.}
The full relation representation achieves the lowest source Meta-test Macro, Worst, and Pooled EER among the broad input controls (Figure~\ref{fig:relation}(a)).
Its Macro EER of 0.069\% is lower than the near-parameter-matched L21-only control (0.119\%) and all single-term-removal variants (0.082--0.114\%; Appendix~\ref{app:input-controls}).
These ablations support retaining both layer summaries and all three comparison terms.

\subsection{Complementarity with the Base Detector}
\label{sec:results-complementarity}

The XLSR-SLS results provide direct evidence of score complementarity: fusion achieves 6.462\% External-5 EER, compared with 7.158\% for the original detector and 11.333\% for Relation alone (Table~\ref{tab:relation_controls}).
For XLSR-Conformer-TCM, fusion likewise improves on the individual scores (6.316\% versus 6.849\% and 6.852\%), while XLSR-AASIST reaches 5.972\% after fusion (Appendix~\ref{app:backend-fusion}).
This indicates that relation-based evidence can improve the combined decision even when the auxiliary score is not stronger on its own.

On the source Meta-test, logistic fusion reduces Macro EER to 0.0690\%, compared with 0.0968\% for Relation alone and 0.0747\% for equal-weight fusion (Figure~\ref{fig:relation}(b)).
The external results use the fusion configuration selected on source data.

\subsection{Geometry-Guided Layer-Pair Selection}
\label{sec:results-pair-selection}

\begin{table}[t]
\centering
\footnotesize
\setlength{\tabcolsep}{3pt}
\begin{tabular}{@{}lcccc@{}}
\toprule
Pair & Gap & $D_G$ & $R_G$ & \shortstack{Macro EER\\(\%)} \\
\midrule
L9--L17  & 8  & 0.45 & 17.66 & 0.1188 \\
L9--L21  & 12 & 0.69 & 17.41 & 0.1062 \\
\textbf{L8--L21} & 13 & 0.85 & 15.26 & \textbf{0.0888} \\
L7--L18  & 11 & 0.72 & 15.22 & 0.0892 \\
L10--L21 & 11 & 0.77 & 14.27 & 0.0965 \\
L7--L19  & 12 & 0.85 & 14.20 & 0.1062 \\
\bottomrule
\end{tabular}
\caption{\textbf{Geometry-guided layer-pair candidates.}
Gap denotes encoder-depth separation, $D_G$ speaker-relative geometry distance, and $R_G$ the candidate-ranking score.
Macro EER is measured on source Meta-dev and determines the selected pair.}
\label{tab:layer_analysis}
\end{table}

Geometry-guided ranking yields six candidates, and detection-based selection chooses L8--L21, which achieves the lowest Meta-dev Macro EER (0.0888\%; Table~\ref{tab:layer_analysis}).
With six candidates, the selected configuration achieves 0.069\% Meta-test Macro EER, compared with 0.078\% for depth-gap-only ranking and 0.103\% for geometry-only ranking.
Random shortlists average 0.076\%, and exhaustive selection achieves 0.071\% (Appendix~\ref{app:candidate-alternatives}).
These results support a compact geometry-guided shortlist followed by detection-based selection.

\subsection{Robustness to Source Partitioning and Initialization}
\label{sec:results-robustness}

\begin{table}[t]
\centering
\footnotesize
\setlength{\tabcolsep}{3pt}
\begin{tabular}{@{}lccrr@{}}
\toprule
Protocol & Pair & $C$ & SLS & + \ours{} \\
\midrule
Utterance-disjoint & L7--L18 & $0.001$ & 0.170 & \textbf{0.101} \\
Speaker-disjoint & L8--L21 & $0.001$ & 0.107 & \textbf{0.045} \\
\bottomrule
\end{tabular}
\caption{\textbf{Source Meta-test results under stricter partitions} (Macro EER, \%).
Each pair and fusion regularization setting is selected independently on Meta-dev.}
\label{tab:strict_main}
\end{table}

\ours{} improves source Meta-test Macro EER under both utterance-disjoint (0.170\% to 0.101\%) and speaker-disjoint (0.107\% to 0.045\%) protocols (Table~\ref{tab:strict_main}).
Geometry and configuration selection are repeated for each partition, supporting the improvements under both stricter source protocols.

Across three runs with L8--L21 fixed, Macro EER averages 0.070\% (sample SD: 0.007 percentage points), compared with 0.149\% for XLSR-SLS.
Repeated layer-pair selection chooses L8--L21 in two runs and L9--L17 in the third, where Macro EER reaches 0.092\%.
These findings support the consistency of the source-side improvements across the evaluated partitions and initializations.

\section{Conclusion}
We introduced CLRM, which augments frozen SSL-based detectors by jointly modeling selected layer summaries and explicit cross-layer comparisons.
With identical layer inputs and approximately matched auxiliary-head parameters, relation-based inputs achieve lower five-dataset mean EER than direct concatenation, both independently and after fusion with XLSR-SLS.
The augmentation improves mean EER for all three evaluated XLS-R-based detectors, with a maximum relative reduction of 16.7\%.
These results support explicit cross-layer comparisons as effective inputs for audio deepfake detector augmentation.

\newpage
\section*{Limitations}

The study uses XLS-R-300M representations, ASVspoof 2019 LA for source-side development, and five external evaluation datasets.
Speaker labels are needed for geometry-based candidate construction, but not for inference.
The primary Meta-dev/Meta-test protocol shares bonafide utterances, while additional utterance- and speaker-disjoint controls assess source-side performance with separately constructed partitions.
Reported external EERs are point estimates; initialization and component ablations are evaluated on source data.
The relation-based system adds three prediction heads to the original detector, and its latency and memory requirements remain to be measured.

\bibliography{refs}

@inproceedings{rosello2023conformer,
  title={A conformer-based classifier for variable-length utterance processing in anti-spoofing},
  author={Rosello, Eros and Gomez-Alanis, Alejandro and Gomez, Angel M and Peinado, Antonio},
  booktitle={Proc. Interspeech 2023},
  pages={5281--5285},
  year={2023}
}

@inproceedings{jung2022aasist,
  title={{AASIST}: Audio Anti-Spoofing Using Integrated Spectro-Temporal Graph Attention Networks},
  author={Jung, Jee-weon and Heo, Hee-Soo and Tak, Hemlata and Shim, Hye-jin and Chung, Joon Son and Lee, Bong-Jin and Yu, Ha-Jin and Evans, Nicholas},
  booktitle={2022 IEEE International Conference on Acoustics, Speech and Signal Processing (ICASSP)},
  doi={10.1109/ICASSP43922.2022.9747766},
  pages={6367--6371},
  year={2022},
  organization={IEEE}
}

@inproceedings{truong24b_interspeech,
  title     = {{Temporal-Channel Modeling in Multi-head Self-Attention for Synthetic Speech Detection}},
  author    = {Duc-Tuan Truong and Ruijie Tao and Tuan Nguyen and Hieu-Thi Luong and Kong Aik Lee and Eng Siong Chng},
  year      = {2024},
  booktitle = {{Interspeech 2024}},
  pages     = {537--541},
  doi       = {10.21437/Interspeech.2024-659},
  issn      = {2958-1796},
}

@article{liu2025nes2net,
  title={Nes2net: A lightweight nested architecture for foundation model driven speech anti-spoofing},
  author={Liu, Tianchi and Truong, Duc-Tuan and Das, Rohan Kumar and Lee, Kong Aik and Li, Haizhou},
  journal={IEEE Transactions on Information Forensics and Security},
  volume={20},
  pages={12005--12018},
  year={2025},
  publisher={IEEE}
}

@inproceedings{zhang2024xlsrsls,
  title={Audio deepfake detection with self-supervised xls-r and sls classifier},
  author={Zhang, Qishan and Wen, Shuangbing and Hu, Tao},
  booktitle={Proceedings of the 32nd ACM International Conference on Multimedia},
  pages={6765--6773},
  year={2024}
}

@inproceedings{pan2024attm,
  title     = {{Attentive Merging of Hidden Embeddings from Pre-trained Speech Model for Anti-spoofing Detection}},
  author    = {Zihan Pan and Tianchi Liu and Hardik B. Sailor and Qiongqiong Wang},
  year      = {2024},
  booktitle = {{Interspeech 2024}},
  pages     = {2090--2094},
  doi       = {10.21437/Interspeech.2024-1472},
  issn      = {2958-1796},
}

@inproceedings{elkheir2025layerwise,
  title={Comprehensive layer-wise analysis of ssl models for audio deepfake detection},
  author={El Kheir, Yassine and Samih, Younes and Maharjan, Suraj and Polzehl, Tim and M{\"o}ller, Sebastian},
  booktitle={Findings of the Association for Computational Linguistics: NAACL 2025},
  pages={4070--4082},
  year={2025}
}

@inproceedings{xiao2025decisionfusion,
  title     = {{Layer-Wise Decision Fusion for Fake Audio Detection Using XLS-R}},
  author    = {Yixuan Xiao and Ngoc Thang Vu},
  year      = {2025},
  booktitle = {{Interspeech 2025}},
  pages     = {5618--5622},
  doi       = {10.21437/Interspeech.2025-1543},
  issn      = {2958-1796},
}

@inproceedings{liang2026hiercon,
  title={HierCon: Hierarchical Contrastive Attention for Audio Deepfake Detection},
  author={Liang, Zhili Nicholas and Han, Soyeon Caren and Wang, Qizhou and Leckie, Christopher},
  booktitle={Proceedings of the ACM Web Conference 2026},
  pages={8413--8416},
  year={2026}
}

@article{wang2020asvspoof2019,
  title={ASVspoof 2019: A large-scale public database of synthesized, converted and replayed speech},
  author={Wang, Xin and Yamagishi, Junichi and Todisco, Massimiliano and Delgado, H{\'e}ctor and Nautsch, Andreas and Evans, Nicholas and Sahidullah, Md and Vestman, Ville and Kinnunen, Tomi and Lee, Kong Aik and others},
  journal={Computer Speech \& Language},
  volume={64},
  pages={101114},
  year={2020},
  publisher={Elsevier}
}

@article{liu2023asvspoof2021,
author = {Liu, Xuechen and Wang, Xin and Sahidullah, Md and Patino, Jose and Delgado, H{\'e}ctor and Kinnunen, Tomi and Todisco, Massimiliano and Yamagishi, Junichi and Evans, Nicholas and Nautsch, Andreas and Lee, Kong Aik},
title = {ASVspoof 2021: Towards Spoofed and Deepfake Speech Detection in the Wild},
year = {2023},
issue_date = {2023},
publisher = {IEEE Press},
volume = {31},
issn = {2329-9290},
doi = {10.1109/TASLP.2023.3285283},
journal = {IEEE/ACM Trans. Audio, Speech and Lang. Proc.},
month = jun,
pages = {2507–2522},
numpages = {16}
}

@inproceedings{wang2024asvspoof5,
  title     = {{ASVspoof 5: crowdsourced speech data, deepfakes, and adversarial attacks at scale}},
  author    = {Xin Wang and Héctor Delgado and Hemlata Tak and Jee-weon Jung and Hye-jin Shim and Massimiliano Todisco and Ivan Kukanov and Xuechen Liu and Md Sahidullah and Tomi H. Kinnunen and Nicholas Evans and Kong Aik Lee and Junichi Yamagishi},
  year      = {2024},
  booktitle = {{The Automatic Speaker Verification Spoofing Countermeasures Workshop (ASVspoof 2024)}},
  pages     = {1--8},
  doi       = {10.21437/ASVspoof.2024-1},
}

@inproceedings{reimao2019for,
  title={FoR: A Dataset for Synthetic Speech Detection.},
  author={Reimao, Ricardo and Tzerpos, Vassilios},
  booktitle={SpeD},
  pages={1--10},
  year={2019}
}

@inproceedings{tak2022automatic,
  title     = {{Automatic Speaker Verification Spoofing and Deepfake Detection Using Wav2vec 2.0 and Data Augmentation}},
  author    = {Hemlata Tak and Massimiliano Todisco and Xin Wang and Jee-weon Jung and Junichi Yamagishi and Nicholas Evans},
  year      = {2022},
  booktitle = {{The Speaker and Language Recognition Workshop (Odyssey 2022)}},
  pages     = {112--119},
  doi       = {10.21437/Odyssey.2022-16},
}

@inproceedings{muller2022does,
  title     = {{Does Audio Deepfake Detection Generalize?}},
  author    = {Nicolas Müller and Pavel Czempin and Franziska Diekmann and Adam Froghyar and Konstantin Böttinger},
  year      = {2022},
  booktitle = {{Interspeech 2022}},
  pages     = {2783--2787},
  doi       = {10.21437/Interspeech.2022-108},
  issn      = {2958-1796},
}

@inproceedings{qin2026dadgmoe,
  title     = {Domain-Adaptive Dual-Gating Mixture of Experts for Generalizable Speech Deepfake Detection},
  author    = {Qin, Siqing and Li, Zhe and Lee, Kong Aik and Mak, Man-Wai},
  booktitle = {Proc. Interspeech},
  year      = {2026}
}

@inproceedings{babu2021xls,
  title     = {{XLS-R: Self-supervised Cross-lingual Speech Representation Learning at Scale}},
  author    = {Arun Babu and Changhan Wang and Andros Tjandra and Kushal Lakhotia and Qiantong Xu and Naman Goyal and Kritika Singh and Patrick {von Platen} and Yatharth Saraf and Juan Pino and Alexei Baevski and Alexis Conneau and Michael Auli},
  year      = {2022},
  booktitle = {{Interspeech 2022}},
  pages     = {2278--2282},
  doi       = {10.21437/Interspeech.2022-143},
  issn      = {2958-1796},
}

@inproceedings{li2024crossdomain,
  title     = {Cross-Domain Audio Deepfake Detection: Dataset and Analysis},
  author    = {Li, Yuang and Zhang, Min and Ren, Mengxin and Qiao, Xiaosong and Ma, Miaomiao and Wei, Daimeng and Yang, Hao},
  editor    = {Al-Onaizan, Yaser and Bansal, Mohit and Chen, Yun-Nung},
  booktitle = {Proceedings of the 2024 Conference on Empirical Methods in Natural Language Processing},
  month     = nov,
  year      = {2024},
  address   = {Miami, Florida, USA},
  publisher = {Association for Computational Linguistics},
  pages     = {4977--4983},
  doi       = {10.18653/v1/2024.emnlp-main.286},
  url       = {https://aclanthology.org/2024.emnlp-main.286/}
}

@inproceedings{elkheir2025bicrossmamba,
  title = {{BiCrossMamba-ST}: Speech Deepfake Detection with Bidirectional {Mamba} Spectro-Temporal Cross-Attention},
  author = {Yassine {El Kheir} and Tim Polzehl and Sebastian M{\"o}ller},
  year = {2025},
  booktitle = {Interspeech 2025},
  pages = {2235--2239},
  doi = {10.21437/Interspeech.2025-527}
}

@inproceedings{reimers-gurevych-2019-sentence,
  title = {Sentence-{BERT}: Sentence Embeddings using Siamese {BERT}-Networks},
  author = {Reimers, Nils and Gurevych, Iryna},
  booktitle = {Proceedings of the 2019 Conference on Empirical Methods in Natural Language Processing and the 9th International Joint Conference on Natural Language Processing (EMNLP-IJCNLP)},
  year = {2019},
  pages = {3982--3992},
  publisher = {Association for Computational Linguistics},
  doi = {10.18653/v1/D19-1410},
  url = {https://aclanthology.org/D19-1410/}
}

@inproceedings{huang-etal-2025-speechfake,
  title = {{S}peech{F}ake: A Large-Scale Multilingual Speech Deepfake Dataset Incorporating Cutting-Edge Generation Methods},
  author = {Huang, Wen and Gu, Yanmei and Wang, Zhiming and Zhu, Huijia and Qian, Yanmin},
  booktitle = {Proceedings of the 63rd Annual Meeting of the Association for Computational Linguistics (Volume 1: Long Papers)},
  month = {July},
  year = {2025},
  address = {Vienna, Austria},
  publisher = {Association for Computational Linguistics},
  pages = {9985--9998},
  doi = {10.18653/v1/2025.acl-long.493},
  url = {https://aclanthology.org/2025.acl-long.493/}
}

@inproceedings{huang-etal-2026-data,
  title = {A Data-Centric Approach to Generalizable Speech Deepfake Detection},
  author = {Huang, Wen and Mao, Yuchen and Qian, Yanmin},
  booktitle = {Proceedings of the 64th Annual Meeting of the {A}ssociation for {C}omputational {L}inguistics (Volume 1: Long Papers)},
  month = {July},
  year = {2026},
  address = {San Diego, California, United States},
  publisher = {Association for Computational Linguistics},
  pages = {17520--17539},
  doi = {10.18653/v1/2026.acl-long.796},
  url = {https://aclanthology.org/2026.acl-long.796/}
}

@inproceedings{chou-etal-2026-iclad,
  title = {{ICLAD}: In-Context Learning with Comparison-Guidance for Audio Deepfake Detection},
  author = {Chou, Benjamin Shiue-Hal and Zhu, Yi and Koppisetti, Surya},
  booktitle = {Findings of the {A}ssociation for {C}omputational {L}inguistics: {ACL} 2026},
  month = {July},
  year = {2026},
  address = {San Diego, California, United States},
  publisher = {Association for Computational Linguistics},
  pages = {9242--9256},
  doi = {10.18653/v1/2026.findings-acl.450},
  url = {https://aclanthology.org/2026.findings-acl.450/}
}
\clearpage

\appendix

\section{Speaker-Relative Geometry and Candidate Search}
\label{app:geometry}

We describe the speaker-relative statistics used for candidate construction and report alternative layer-pair searches and a post-hoc cosine analysis.

\subsection{Speaker-Relative Statistics}
\label{app:geometry-statistics}

Let $\mathcal{B}_k$ be the set of bona fide reference utterances from speaker $k$, with $K$ speakers and $\mathcal{B}=\bigcup_{k=1}^{K}\mathcal{B}_k$.
For utterance $x_i$, we construct a geometry embedding at layer $l$ by temporal mean pooling followed by $\ell_2$ normalization:
\begin{equation}
\begin{aligned}
\bar{\mathbf{h}}_i^{(l)}
&=\frac{1}{T_i^{(l)}}\sum_{t=1}^{T_i^{(l)}}\mathbf{H}_{i,t}^{(l)},\\
\mathbf{h}_i^{(l)}
&=\frac{\bar{\mathbf{h}}_i^{(l)}}{\|\bar{\mathbf{h}}_i^{(l)}\|_2}.
\end{aligned}
\label{eq:app-geometry-embedding}
\end{equation}
For speaker $k$, the centroid and the leave-one-out centroid for an utterance $i\in\mathcal{B}_k$ are
\begin{equation}
\begin{aligned}
\mathbf{c}_k^{(l)}
&=\frac{1}{|\mathcal{B}_k|}\sum_{j\in\mathcal{B}_k}\mathbf{h}_j^{(l)},\\
\mathbf{c}_{k,-i}^{(l)}
&=\frac{1}{|\mathcal{B}_k|-1}\sum_{\substack{j\in\mathcal{B}_k\\j\ne i}}\mathbf{h}_j^{(l)}.
\end{aligned}
\label{eq:app-centroids}
\end{equation}
We compute the cosine similarity to the correct speaker and each competing speaker $k'\ne k$ as
\begin{equation}
\begin{aligned}
s_{i,k}^{(l)}
&=\cos\!\left(\mathbf{h}_i^{(l)},\mathbf{c}_{k,-i}^{(l)}\right),\\
s_{i,k'}^{(l)}
&=\cos\!\left(\mathbf{h}_i^{(l)},\mathbf{c}_{k'}^{(l)}\right).
\end{aligned}
\label{eq:app-speaker-similarities}
\end{equation}
Let $r_i^{(l)}$ be the rank of the correct speaker among the $K$ scores in descending order, and let $a_i^{(l)}=\mathbb{I}[r_i^{(l)}=1]$.
The margins relative to the strongest and average competing speakers are
\begin{equation}
\begin{aligned}
m_{i,\max}^{(l)}
&=s_{i,k}^{(l)}-\max_{k'\ne k}s_{i,k'}^{(l)},\\
m_{i,\mathrm{mean}}^{(l)}
&=s_{i,k}^{(l)}-\frac{1}{K-1}\sum_{k'\ne k}s_{i,k'}^{(l)}.
\end{aligned}
\label{eq:app-speaker-margins}
\end{equation}
Averaging over the reference utterances gives four layer-level statistics:
\begin{equation}
\begin{aligned}
R_l&=\frac{1}{|\mathcal{B}|}\sum_{i\in\mathcal{B}}r_i^{(l)},\\
A_l&=\frac{1}{|\mathcal{B}|}\sum_{i\in\mathcal{B}}a_i^{(l)},\\
M_l^{\max}&=\frac{1}{|\mathcal{B}|}\sum_{i\in\mathcal{B}}m_{i,\max}^{(l)},\\
M_l^{\mathrm{mean}}&=\frac{1}{|\mathcal{B}|}\sum_{i\in\mathcal{B}}m_{i,\mathrm{mean}}^{(l)}.
\end{aligned}
\label{eq:app-geometry-aggregation}
\end{equation}
The four statistics form $\mathbf{g}_l=[R_l;A_l;M_l^{\max};M_l^{\mathrm{mean}}]$.
These mean-pooled, $\ell_2$-normalized geometry embeddings are distinct from the mean-and-standard-deviation summaries used to construct relation inputs in Eq.~\ref{eq:cross-layer-relation}.

\subsection{Geometry Normalization and Search Domain}
\label{app:geometry-search}

Each coordinate of $\mathbf{g}_l$ is standardized across candidate layers using its population mean and standard deviation.
We consider layers L1--L21 with a minimum separation of eight layers, yielding 91 eligible pairs.
The ranking score $R_G(p,q)$ in Eq.~\ref{eq:geometry-ranking} combines depth separation with similarity of the standardized geometry vectors.
The six highest-ranked pairs form the shortlist for detection-based selection on Meta-dev.
The geometry statistics determine the candidate shortlist, while the relation classifier uses the representation in Eq.~\ref{eq:cross-layer-relation}.

\subsection{Alternative Candidate-Selection Strategies}
\label{app:candidate-alternatives}

We compare the geometry-guided six-pair shortlist with ten random six-pair shortlists and an exhaustive search over all 91 eligible pairs.
All strategies select the layer pair and fusion regularization on Meta-dev by fused Macro EER and evaluate the selection on Meta-test.
Random shortlists are sampled without replacement using NumPy's default random generator (seed 20261001), with eligible pairs ordered by their lower and upper layer indices.

\begin{table}[t]
\centering
\small
\setlength{\tabcolsep}{3.5pt}
\caption{\textbf{Alternative candidate searches} (source Macro EER, \%; lower is better).
Each random shortlist contains six pairs; exhaustive search considers all 91 eligible pairs.
The pair and fusion regularization are selected on Meta-dev and evaluated on Meta-test.}
\label{tab:app-search}
\begin{tabular}{@{}llrr@{}}
\toprule
Search & Selected pair & Meta-dev & Meta-test \\
\midrule
Random 00 & L2--L16 & 0.068 & 0.085 \\
Random 01 & L3--L11 & 0.064 & 0.071 \\
Random 02 & L2--L11 & 0.069 & 0.064 \\
Random 03 & L3--L17 & 0.069 & 0.064 \\
Random 04 & L2--L17 & 0.075 & 0.085 \\
Random 05 & L10--L21 & 0.097 & 0.075 \\
Random 06 & L2--L21 & 0.072 & 0.078 \\
Random 07 & L2--L12 & 0.065 & 0.081 \\
Random 08 & L2--L16 & 0.068 & 0.085 \\
Random 09 & L2--L19 & 0.091 & 0.076 \\
\midrule
Exhaustive (91) & L3--L11 & 0.064 & 0.071 \\
\bottomrule
\end{tabular}
\end{table}

The random shortlists yield a mean Meta-test Macro EER of 0.076\% (sample standard deviation: 0.008 percentage points).
The geometry-guided configuration achieves 0.069\%, while the configuration selected by exhaustive search achieves 0.071\% (Table~\ref{tab:app-search}).

\subsection{Attack-Wise Cross-Layer Cosine Analysis}
\label{app:cosine-analysis}

For the selected L8--L21 pair, we measure $\Delta_{\cos}=\bar{c}_{\mathrm{bona\ fide}}-\bar{c}_{\mathrm{spoof}}$, where each term is the corresponding class-mean cosine similarity between standardized layer summaries on source Meta-test.
Table~\ref{tab:app-attack-cosine} reports the attack-wise results.

\begin{table}[t]
\centering
\scriptsize
\setlength{\tabcolsep}{3.5pt}
\caption{Attack-wise $\Delta_{\cos}$ for L8--L21 on source Meta-test ($10^{-2}$).}
\label{tab:app-attack-cosine}
\begin{tabular}{@{}lrrrrrr@{}}
\toprule
Attack & A08 & A10 & A12 & A14 & A16 & A18 \\
\midrule
$\Delta_{\cos}$ & 6.86 & 3.68 & 1.81 & 4.74 & $-9.45$ & 5.00 \\
\bottomrule
\end{tabular}
\end{table}

This statistic is used only for post-hoc analysis and does not enter layer-pair selection or score fusion.

\section{Source Protocol and Implementation Details}
\label{app:source-details}

\subsection{Construction of Disjoint Source Protocols}
\label{app:protocol-construction}

The primary source protocol assigns attacks A07, A09, A11, A13, A15, A17, and A19 to Meta-dev, and A08, A10, A12, A14, A16, and A18 to Meta-test.
Both partitions contain the same 7,355 bona fide utterances from 67 speakers, which also provide the primary geometry reference.
The utterance-disjoint and speaker-disjoint controls retain the attack assignments while separating bona fide utterances or speakers between Meta-dev and Meta-test.
In both controls, the geometry reference consists of 2,580 bona fide training utterances from 20 speakers, and candidate construction and configuration selection are repeated independently.

Both disjoint protocols use seed 20261001.
For the utterance-disjoint protocol, utterance IDs are sorted within each speaker by the first 16 hexadecimal digits of a Secure Hash Algorithm 256 (SHA-256) hash of the seed, a null separator, and the utterance ID.
Let $n_k$ denote the number of bona fide utterances available for speaker $k$. The first $\lceil n_k/2\rceil$ utterances are assigned to Meta-dev and the remainder to Meta-test, giving 3,687 and 3,668 bona fide utterances, respectively.
For the speaker-disjoint protocol, speakers are sorted by decreasing trial count with seed-dependent hash tie-breaking, then greedily assigned to the partition with fewer trials.
This gives 30 Meta-dev speakers (3,330 bona fide trials) and 37 Meta-test speakers (4,025 bona fide trials).
Table~\ref{tab:source-protocols} provides the trial counts and overlap statistics.

\subsection{Relation-Head Fitting and Score Fusion}
\label{app:fitting}

For each candidate pair, three independently initialized relation heads are trained on the official source training split, with checkpoints selected on the official source development split.
Their raw scores are averaged before score orientation, normalization, and fusion.
Score orientation refers to the choice of score sign.
For the primary XLSR-SLS experiments, this sign is selected on Meta-dev by minimizing $0.50\,\mathrm{Pooled}+0.35\,\mathrm{Macro}+0.15\,\mathrm{Worst}$ EER.
Worst and Pooled EER are successive tie-breakers, and the positive orientation is retained if all criteria tie.

Each oriented score stream is standardized using the official source development mean $\mu$ and population standard deviation $\sigma$:
\begin{equation}
\tilde{s}(x)=\frac{s(x)-\mu}{\max(\sigma,10^{-8})}.
\label{eq:app-score-normalization}
\end{equation}
The normalized detector and relation scores are fused using logistic regression with an intercept, $\ell_2$ regularization, and no class reweighting.
We use the limited-memory Broyden--Fletcher--Goldfarb--Shanno (L-BFGS) solver with at most 5,000 iterations and tolerance $10^{-10}$.
The inverse regularization strength is selected jointly with the layer pair on Meta-dev from $C\in\{10^{-4},10^{-3},10^{-2},0.1,1,10\}$ using fused Macro EER.
The selected orientation, normalization statistics, layer pair, and fusion configuration are fixed for Meta-test and external evaluation.

\subsection{Auxiliary Inputs and Operator Controls}
\label{app:input-controls}

Table~\ref{tab:app-input-budgets} summarizes the input dimensions, hidden widths, and trainable parameters of the auxiliary heads.
Parameter counts refer to one auxiliary head and exclude the frozen SSL encoder, original detection head, and fusion coefficients.
Each configuration uses three independently trained heads.
Concat-256 matches the hidden width of Relation, while Concat-512 approximately matches its per-head parameter count.
Single-layer and relation-term removal controls are trained independently using the same source-side fitting procedure.

\begin{table}[t]
\centering
\scriptsize
\setlength{\tabcolsep}{2.4pt}
\caption{\textbf{Auxiliary input dimensions and trainable parameters per head.}
Each configuration uses three independently trained heads.}
\label{tab:app-input-budgets}
\begin{tabular}{@{}lrrr@{}}
\toprule
Input & Dim. & Width & Params/head \\
\midrule
Relation & 8,193 & 256 & 2,097,921 \\
Concat-256 & 4,096 & 256 & 1,049,089 \\
Concat-512 & 4,096 & 512 & 2,098,177 \\
L21-256 & 2,048 & 256 & 524,801 \\
L21-1023 & 2,048 & 1,023 & 2,097,151 \\
w/o abs. difference & 6,145 & 256 & 1,573,633 \\
w/o product & 6,145 & 256 & 1,573,633 \\
w/o cosine & 8,192 & 256 & 2,097,665 \\
\bottomrule
\end{tabular}
\end{table}

Table~\ref{tab:app-input-ablations} reports fused Macro EER on source Meta-test for the single-layer and relation-component controls.
Each configuration uses independently fitted heads and Meta-dev-selected fusion regularization; the full Relation input is reported in Table~\ref{tab:relation_controls}.

\begin{table}[t]
\centering
\small
\setlength{\tabcolsep}{3.5pt}
\caption{\textbf{Additional representation controls on source Meta-test} (fused Macro EER, \%; lower is better).
$C$ is independently selected on Meta-dev.}
\label{tab:app-input-ablations}
\begin{tabular}{@{}lrr@{}}
\toprule
Auxiliary input & Selected $C$ & Macro EER \\
\midrule
L21-256 & $0.1$ & 0.121 \\
L21-1023 & $10$ & 0.119 \\
w/o absolute difference & $10^{-4}$ & 0.114 \\
w/o product & $1$ & 0.110 \\
w/o cosine & $10^{-3}$ & 0.082 \\
\bottomrule
\end{tabular}
\end{table}

The distance-statistics input in Figure~\ref{fig:relation} comprises eight scalar features: cosine similarity, mean squared difference, mean absolute difference, maximum absolute difference, the 90th percentile of absolute difference, log ratio of vector norms, and mean absolute difference for the temporal-mean and temporal-standard-deviation components separately.
Cosine similarity and norm ratios use $\epsilon=10^{-8}$ for numerical stability.
This control uses a head with hidden width 32 and is independently trained from the capacity-controlled concatenation configurations.

\subsection{Optimization, Preprocessing, and Ensemble Accounting}
\label{app:optimization}

Audio is converted to mono at 16~kHz and represented by 64,600 samples; longer utterances are truncated and shorter ones are repeated to this length.
Temporal standard deviations are floored at $10^{-6}$, and coordinate-wise feature standardization uses a denominator floor of $10^{-5}$.
The XLS-R-300M encoder used by the existing detector has hidden dimension $d=1024$ and remains frozen.

Each relation head follows a Linear--GELU--Dropout--Linear architecture with hidden width 256 and dropout 0.20.
Training uses AdamW with learning rate $2\times10^{-4}$, weight decay $10^{-4}$, batch size 2,048, gradient clipping at 5.0, and at most 15 epochs.
Early stopping uses official-development EER with patience four.
The primary representation experiments use weighted binary cross-entropy with spoof encoded as class 1 and positive-class weight $n_0/\max(n_1,1)$, where $n_0$ and $n_1$ denote training-set class counts.

The nominal seeds are 1111, 1337, and 2026, with a recorded method offset of $+300$, giving effective seeds 1411, 1637, and 2326.
For each configuration, the three raw head outputs are averaged before score orientation, normalization, and fusion.
Reported ensemble EERs are computed from the averaged score stream.

\subsection{Initialization Sensitivity}
\label{app:initialization}

With L8--L21 and $C=10^{-4}$ fixed, the recorded initialization check gives source Meta-test Macro EERs of 0.063\%, 0.072\%, and 0.075\%.
The recorded summary reports a mean of 0.070\% and a sample standard deviation of 0.007 percentage points, compared with 0.149\% for the XLSR-SLS baseline.
When layer-pair selection is repeated, L8--L21 is selected in two runs and L9--L17 in the third; the latter obtains 0.092\% Macro EER on Meta-test.
These are source-side evaluations, separate from the reported external-dataset EERs.

\section{Additional Detector Fusion Results}
\label{app:backend-fusion}

Table~\ref{tab:app-backend-fusion} reports relation-only scoring, equal-weight fusion of source-normalized scores, and source-selected logistic fusion for XLSR-Conformer-TCM and XLSR-AASIST.
Both detectors use the layer indices selected with XLSR-SLS and detector-specific relation heads and fusion coefficients.
The main results use the logistic-fusion configuration selected on source data.

\begin{table}[t]
\centering
\small
\setlength{\tabcolsep}{4pt}
\caption{\textbf{Additional detector fusion results} (External-5 mean EER, \%; lower is better).
Logistic denotes the source-selected configuration used in the main results.}
\label{tab:app-backend-fusion}
\begin{tabular}{@{}lrrrr@{}}
\toprule
Detector & Base & Relation & Equal & Logistic \\
\midrule
TCM & 6.849 & 6.852 & 6.295 & 6.316 \\
AASIST & 7.170 & 5.980 & 5.815 & 5.972 \\
\bottomrule
\end{tabular}
\end{table}

\section{External Evaluation and Reproducibility}
\label{app:reproducibility}

\subsection{External Evaluation Populations}
\label{app:external-populations}

Table~\ref{tab:app-external-populations} reports the aligned score-record populations for the primary XLSR-SLS representation comparisons.
External-5 is the unweighted mean of the five dataset-level EERs, rather than an EER computed by pooling their trials.

\begin{table}[h]
\centering
\small
\setlength{\tabcolsep}{4pt}
\caption{\textbf{Aligned external score-record populations} for the XLSR-SLS relation comparison.}
\label{tab:app-external-populations}
\begin{tabular}{@{}lr@{}}
\toprule
External test set & Utterances \\
\midrule
ASVspoof 2021 LA & 181,566 \\
ASVspoof 2021 DF & 611,829 \\
ASVspoof 5 Eval & 680,774 \\
Normalized Fake-or-Real test & 4,634 \\
In-the-Wild & 31,212 \\
\bottomrule
\end{tabular}
\end{table}

For source and generic internal evaluation, we select the operating point that minimizes the absolute difference between the false-positive rate (FPR) and false-negative rate (FNR).
At this point, EER is reported as $100(\mathrm{FPR}+\mathrm{FNR})/2$.
For ASVspoof 2021 LA and DF, we retain the EER values from the official evaluators.

\subsection{Result Sources and Artifact Coverage}
\label{app:result-provenance}

Table~\ref{tab:app-provenance} summarizes the evaluation records and supporting artifacts in the assembled result archive.
Reference systems are represented by EER summaries, while the XLSR-SLS relation comparison includes aligned score streams and source-side fusion records.
The assembled archive contains source configurations and per-dataset evaluations for XLSR-Conformer-TCM and XLSR-AASIST, but not their auxiliary checkpoint-to-score mappings.

\begin{table}[h]
\centering
\footnotesize
\setlength{\tabcolsep}{3pt}
\caption{\textbf{Evaluation records and supporting materials.}
Coverage refers to the assembled result archive.}
\label{tab:app-provenance}
\begin{tabular}{@{}p{.40\columnwidth}p{.51\columnwidth}@{}}
\toprule
System & Available records \\
\midrule
XLSR-Conformer & EER summary \\
Nes2Net-X & EER summary \\
XLSR-DADGMoE-AASIST & EER summary \\
BiCrossMamba-ST & EER summary \\
XLSR-SLS + \ours{} & Aligned scores; fusion settings; source controls \\
XLSR-Conformer-TCM + \ours{} & Dataset results; source configurations \\
XLSR-AASIST + \ours{} & Dataset results; source configurations \\
\bottomrule
\end{tabular}
\end{table}

\subsection{Score Replay and Verification Scope}
\label{app:score-replay}

The XLSR-SLS L8--L21 records include aligned base, relation, and fused scores for all five external test sets.
Reapplying the saved normalization and fusion parameters reproduces the archived relation-based scores with a maximum absolute difference of $6.4\times10^{-7}$, without refitting.
Relation and Concat use the same L8--L21 feature caches and standardization statistics, with independently trained heads and source-fitted fusion coefficients.
Saved source score streams and fusion configurations also support replay of the corresponding source-side control metrics.
For XLSR-Conformer-TCM and XLSR-AASIST, the retained per-dataset evaluation values allow recomputation of External-5 from unrounded results.

\end{document}